\documentclass{article}
\usepackage{spconf,amsmath,amssymb,graphicx}
\usepackage{multirow}
\usepackage{color}
\usepackage{subcaption}
\usepackage{url}

\title{Multimodal Confidence Modeling in Audio--Visual Quality Assessment}
\name{Mayesha Maliha R. Mithila and Mylene C.Q. Farias}
\address{Texas State University\\
Department of Computer Science}
\usepackage{etoolbox}
\usepackage{setspace}

\apptocmd{\thebibliography}{%
  \setlength{\itemsep}{0pt}%
  \setlength{\parskip}{0pt}%
  \setstretch{0.98}%
}{}{}

\begin{document}
%\ninept
%
\maketitle

\footnotetext{© 2026 IEEE. Personal use of this material is permitted. Permission from IEEE must be obtained for all other uses, in any current or future media, including reprinting/republishing this material for advertising or promotional purposes, creating new collective works, for resale or redistribution to servers or lists, or reuse of any copyrighted component of this work in other works.}
\begin{abstract}
Audio-visual quality assessment (AVQA) is essential for streaming, teleconferencing, and immersive media. In realistic streaming scenarios, distortions are often asymmetric, where one modality may be severely degraded while the other remains clean. Still, most contemporary AVQA metrics treat audio and video as equally reliable, causing confidence-unaware fusion to emphasize unreliable signals. This paper proposes MCM-AVQA, a multimodal confidence-aware AVQA framework that explicitly estimates modality-specific confidence and injects it into a dedicated Audio–Visual Mixer for cross-modal attention. The Audio–Visual Mixer utilizes frame-level, confidence-guided channel attention to gate fusion, modulating feature interaction between modalities so that high-confidence streams dominate while unreliable inputs are suppressed, preserving temporal degradation patterns. A multi-head visual confidence estimator turns frame-level artifact probabilities into temporally smoothed, clip-level visual confidence scores, while an audio confidence module derives confidence from speech-quality cues without requiring a clean reference. Experiments on multiple AVQA benchmarks show that MCM-AVQA, and specifically its confidence-guided Audio–Visual Mixer, improve correlation with human mean opinion scores and yield more interpretable behavior under real-world asymmetric audio-visual distortions.

\end{abstract}

\section{Introduction}
\label{sec:intro}

Audio-visual quality assessment (AVQA) is essential for streaming, teleconferencing, and immersive media because it allows for adaptive streaming and large-scale quality monitoring without human intervention ~\cite{akhtar2017audio}. Under realistic operating conditions, however, audiovisual signals are often subject to asymmetric degradations. In such cases, the video stream may exhibit spatial artifacts, such as blocking or blurring, as well as temporal distortions, including frame freezing, while the accompanying audio signal remains undistorted. In contrast, the audio may be degraded by coding distortion or packet loss, while visual quality is mostly preserved~\cite{min2020livesjtu}. Psychophysical evidence shows that humans weight such asymmetrically degraded signals by reliability rather than using a weakest-link rule~\cite{rohe2018reliability}.

Early AVQA approaches often fused separately predicted audio and visual quality scores using simple pooling functions (e.g., linear or Minkowski pooling), offering limited modeling of cross-modal interaction. Although recent audio–visual quality assessment (AVQA) methods leverage content-adaptive attention to improve quality prediction~\cite{cao2023attention}, many AVQA algorithms still treat audio and visual modalities as equally reliable and do not explicitly model modality confidence. Subsequent deep learning-based AVQA studies have primarily focused on learning joint audio-visual representations from pre-extracted unimodal features. NAViDAd \cite{martinez2019navidad} learns a shared latent embedding space via a two-stage autoencoder framework; however, it lacks explicit cross-modal interaction mechanisms and does not enforce modality reliability under asymmetric degradations. Attention-guided AVQA architectures \cite{cao2023attention} integrate visual saliency mechanisms with late fusion, where attention weights are learned implicitly, yet without explicitly modeling modality confidence.

\begin{figure*}[h!]
    \centering
    \includegraphics[width=1.0\textwidth]{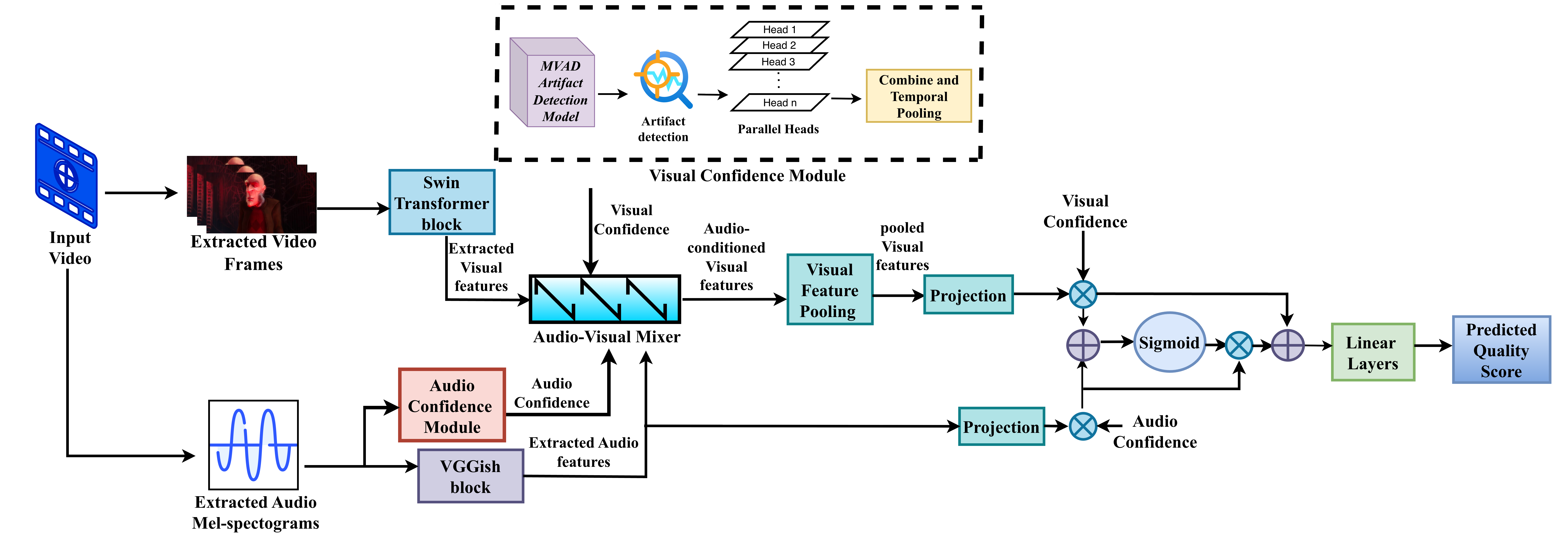}
    \caption{Overall architecture of MCM-AVQA. Swin and VGGish encode the video and audio streams, while specific modules estimate visual and audio confidences. The confidence-aware Audio-Visual Mixer then modulates cross-modal attention before predicting the overall audio-visual quality.}
    \label{fig:pipeline}
\end{figure*}

More recent transformer-based models, such as AVSegFormer~\cite{gao2024avsegformer} for audio-visual segmentation and MMAudio~\cite{cheng2025mmaudio} for video-to-audio generation using joint transformers and synchronization modules, provide effective audio-visual fusion for recognition and generation tasks. Recent quality assessment approaches have similarly leveraged temporal modeling and dynamic attention mechanisms~\cite{mithila2026convolutions} as well as multimodal fusion with temporal transformers~\cite{chakraborty2025mt} to improve perceptual alignment. Nonetheless, these architectures are not designed to modulate fusion as a function of modality asymmetry, a capability that is essentiaal for quality assessment. Attend-Fusion \cite{awan2024attend}, an attention-based late fusion method, enhances video categorization performance but does not explicitly encode modality confidence. Overall, this gap indicates that although attention mechanisms facilitate more informative feature selection, they do not explicitly quantify the degree of trust that should be assigned to each modality in the presence of distortion.

Recent research efforts have increasingly focused on unified multimodal quality assessment. UNQA~\cite{cao2025unqa} employs shared feature extractors alongside modality-specific regressors to perform blind quality prediction for audio, image, video, and audio-visual content. However, it treats all modalities equally and does not explicitly model or quantify confidence. Beyond AVQA, confidence-weighted fusion has been shown to enhance robustness in other audio-visual tasks, such as biometric identification~\cite{alam2015confidence}, by adapting modality contributions typically through confidence-weighted \emph{decision-level} fusion. These prior approaches, however, are predominantly tailored to classification problems with discrete labels rather than to continuous perceptual quality estimation in practical streaming scenarios, where audio and video may contribute in highly asymmetric ways to the overall perceived quality.

In this paper, we propose \textbf{MCM-AVQA}, a \emph{multimodal confidence modeling} framework for AVQA. In contrast to purely channel-wise feature fusion without explicit confidence inputs, MCM-AVQA explicitly estimates modality-specific confidence scores and injects them into a dedicated fusion mechanism. The main contributions of this work are:
\begin{itemize}
\item A \emph{confidence-aware Audio--Visual Mixer} that employs audio and visual confidences to gate cross-modal fusion at the feature level, rather than only at the decision level, via frame-wise confidence-augmented channel attention, so that high-confidence streams dominate, while unreliable inputs are suppressed.
\item A confidence-aware AVQA framework including a \emph{visual confidence module} that maps a multiple visual artifact detector's (MVAD) \cite{feng2025mvad} artifact probabilities to temporally smoothed clip-level visual confidence, and a \emph{audio confidence module} that derives confidence from no-reference SCOREQ-based speech-quality cues~\cite{ragano2024scoreq}.
% \item We demonstrate that the proposed confidence-aware attention improves robustness and yields more stable and interpretable cross-modal behavior under asymmetric distortion than confidence-unaware AVQA baselines.
\item We conduct extensive evaluations and ablations showing how confidence-aware fusion addresses modality contributions and improves robustness under asymmetric distortions.
\end{itemize}

\section{Methodology}
\label{sec:method}

MCM-AVQA incorporates modality-specific confidence into the cross-modal attention and fusion method. Unlike task-specific architectures such as AVSegFormer~\cite{gao2024avsegformer} (segmentation with symmetric channel-attention mixers) or MMAudio~\cite{cheng2025mmaudio} (video-to-audio generation with joint-attention transformers), our approach uses cross-modal attention with explicit confidence estimation. Per-frame channel (feature) gating is done in the Audio-Visual Mixer (AVM) and clip-level weighting is done in the global fusion module. Figure~\ref{fig:pipeline} provides an overview of the complete system. 

In a video clip of $T$ frames, each frame is first processed through a Swin Transformer backbone \cite{liu2021swin} to obtain hierarchical visual feature maps that represent local and global spatial organization. The final-stage features are combined and projected into a compact latent space, which is used as input for the visual confidence estimator and the fusion modules, resulting in a shared representation across all frames. For acquiring visual confidence, the pretrained Multiple Visual Artifact Detector (MVAD) module~\cite{feng2025mvad} analyzes each frame and provides probabilities for the types of artifacts $K=10$ (blocking, blurring, ringing, color bleeding, flickering, banding, mosquito noise, frame freezing, frame dropping and jerkiness). These probabilities are collected into a matrix $A \in \mathbf{R}^{T \times K}$, where $T$ is the number of frames and $A_{t,k}$ denotes the probability of artifact $k$ in frame $t$. 
To capture temporal artifact dynamics, we apply depthwise 1D convolution along the temporal axis, smoothing each artifact channel independently to obtain the smoothed matrix $X$:
\begin{equation}
X = \text{TemporalConv1D}(A) \in \mathbf{R}^{T \times K} .
\end{equation}

\begin{figure}[tb!]
    \centering
    \includegraphics[width=0.45\textwidth]{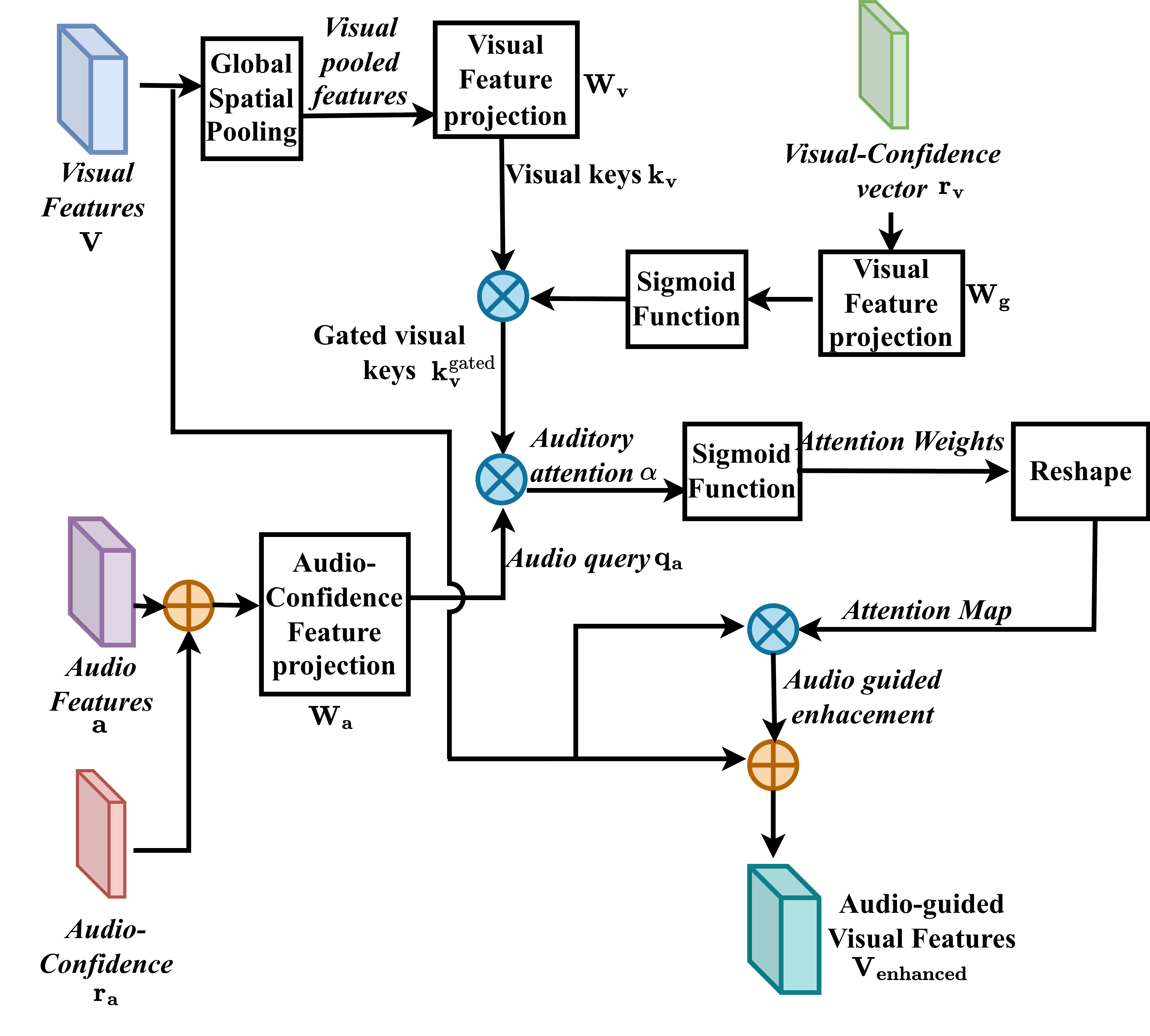}
    \caption{The Confidence-aware Audio–Visual Mixer uses visual and audio features, together with modality confidence scores, to compute channel-wise auditory attention weights that modulate visual feature maps and produce audio-guided visual representations.}
    \label{fig:pipeline}
\end{figure}

For each frame $t$, the smoothed vector $X_t \in \mathbf{R}^K$ is passed through parallel Multi Layer Perception (MLP) heads followed by a combiner network to produce a scalar frame-level confidence score $r_t \in [0,1]$:
\begin{equation}
r_t = \text{Combiner}(\text{Heads}(X_t)) .
\end{equation}
The final clip-level visual confidence $r_v$ is the temporal average of frame-level scores $r_t$, representing the overall visual trustworthiness.

The audio recording is first transformed into log-mel spectrograms and subsequently encoded using a VGGish encoder \cite{hershey2017cnn}, yielding a clip-level embedding that compactly characterizes the acoustic content. This representation is further refined through lightweight attention layers and is shared between the audio confidence estimation module and the Audio-Visual Mixer, thereby ensuring that both confidence estimation and multimodal fusion operate on an identical set of audio-derived attributes. Leveraging SCOREQ-based speech-quality modeling \cite{ragano2024scoreq}, an audio confidence score $r_a \in [0,1]$ is inferred from no-reference speech-quality predictions. In practice, we obtain a SCOREQ quality score like estimate for each audio clip, apply min--max normalization to map it into $[0,1]$, and use the resulting scalar as $r_a$.

\textbf{Confidence-Aware Audio-Visual Mixer.} Instead of employing a symmetric fusion scheme, the proposed mixer modulates cross-modal interactions through learned confidence estimates and channel attention. Given visual features $V \in \mathbf{R}^{B \times C \times H \times W}$, projected audio features $a \in \mathbf{R}^{B \times d}$, and corresponding confidence scores $r_a \in \mathbf{R}^{B \times 1}$ and $r_v \in \mathbf{R}^{B \times 1}$, for each frame $t$, we construct an audio query by concatenating the audio features with their confidence and mapping the result to the visual feature dimensionality:
\begin{equation}
q_a = W_a[a; r_a] \in \mathbf{R}^{B \times C},
\end{equation}
where $W_a \in \mathbf{R}^{(d+1) \times C}$ is a learnable projection matrix and $[a; r_a]$ denotes concatenation along the feature dimension.

The visual feature tensor $V$ is first subjected to global average pooling and subsequently mapped into the key space via a learned linear projection:
\begin{equation}
k_v = W_v\bigl(\text{GAP}(V)\bigr) \in \mathbb{R}^{B \times C},
\end{equation}
where $W_v \in \mathbb{R}^{C \times C}$ denotes a trainable projection matrix that implements a linear transformation, and $\text{GAP}(\cdot)$ represents the global average pooling operator. Visual confidence scores $r_v$ are used to gate visual keys and obtain gated keys $k_v^{\text{gated}}$, thus attenuating unreliable visual signals:
\begin{equation}
k_v^{\text{gated}} = k_v \odot \sigma\bigl(W_g(r_v)\bigr) \in \mathbb{R}^{B \times C} ,
\end{equation}
where $W_g \in \mathbb{R}^{1 \times C}$ denotes a linear projection that maps the visual confidence $r_v$ from $\mathbb{R}^{B \times 1}$ to $\mathbb{R}^{B \times C}$, $\sigma(\cdot)$ represents the sigmoid activation function, and $\odot$ indicates element-wise (Hadamard) multiplication.

The audio query $q_a$ and the gated visual keys $k_v^{\text{gated}}$ are multiplied element-by-element and passed through the sigmoid to produce channel-by-channel attention weights $\alpha$:
\begin{equation}
\alpha = \sigma(q_a \odot k_v^{\text{gated}}) \in \mathbf{R}^{B \times C} ,
\end{equation}
where $\alpha$ represents the learned attention weights in all $C$ channels for each sample in the batch. These channel-wise weights are reshaped to spatial dimensions and applied as a residual to enhance the visual features, allowing high-confidence modalities to dominate fusion while attenuating unreliable inputs:
\begin{equation}
V_{\text{enhanced}} = V + V \odot \text{Reshape}(\alpha) \in \mathbf{R}^{B \times C \times H \times W},
\end{equation}
where $\text{Reshape}(\alpha)$ transmits $\alpha$ from $(B, C)$ to $(B, C, 1, 1)$ and applies the channel weights uniformly at all spatial positions $(H, W)$.

Finally, clip-level visual and audio features, together with their associated confidence scores, are integrated by a lightweight fusion network that adaptively emphasizes reliable modalities while attenuating the contribution of degraded ones. The resulting fused representation is subsequently fed into a regression head to estimate the overall audio–visual quality score. This design enables modalities with high confidence to dominate the fusion process while suppressing the impact of low-confidence inputs, thereby ensuring robust performance under asymmetric distortion conditions. Specifically, when the audio confidence is high, the audio queries exert a strong influence on the visual keys, whereas, when the visual confidence is high, the gated visual keys predominantly govern the attention computation.

\textbf{Loss functions.} For each training mini-batch of size \(N\), the model outputs a set of predicted quality scores \(\hat{\mathbf{y}} = [\hat{y}_1, \dots, \hat{y}_N]\) corresponding to the ground-truth MOS annotations \(\mathbf{y} = [y_1, \dots, y_N]\). The training objective combines the mean squared error (MSE) between \(\hat{y}_i\) and \(y_i\) with a Pearson-correlation-based loss,
\[
\mathcal{L}_{\mathrm{PCC}} = 1 - \rho(\hat{\mathbf{y}}, \mathbf{y}),
\]
where \(\rho(\hat{\mathbf{y}}, \mathbf{y})\) denotes the Pearson correlation coefficient computed over the \(N\) prediction–label pairs in the mini-batch. The overall training loss is defined as
\[
\mathcal{L} = \mathcal{L}_{\mathrm{MSE}} + \lambda\, \mathcal{L}_{\mathrm{PCC}},
\]
where \(\mathcal{L}_{\mathrm{MSE}}\) is the conventional MSE between \(\hat{\mathbf{y}}\) and \(\mathbf{y}\), and \(\lambda = 0.15\) is a weighting hyperparameter that balances absolute value fidelity against monotonic agreement in ranking.

\section{Experimental Results}
\label{sec:results}

% We evaluate MCM-AVQA on three AVQA datasets: UnB-AV and UnB-AVQ \footnote{\url{https://www.cdvl.org/} (CDVL member access; video IDs: 2980, 2981).}, and LIVE-SJTU \footnote{\url{https://live.ece.utexas.edu/research/avqa/index.html}}. 

We evaluate MCM-AVQA on three AVQA datasets: UnB-AV \cite{martinez2020unbav}, UnB-AVQ\cite{martinez2014full} and LIVE-SJTU\cite{min2020livesjtu}. These databases contain diverse audio-visual content and distortions, each with subjective mean opinion scores (MOS). Performance is measured by the Pearson Linear Correlation Coefficient (PLCC) and Spearman Rank-Order Correlation Coefficient (SROCC), computed after standard logistic regression between objective predictions and MOS normalized to \([0,1]\) \cite{seshadrinathan2010study}. For visual feature extraction, the Swin-Small architecture is used to derive spatiotemporal representations from some uniformly sampled frames per video (e.g. 8), while VGGish is used to obtain audio embeddings from log-mel spectrograms \cite{liu2021swin, hershey2017cnn}. For each dataset, split ratio of train:validation:test = 70:15:15 is used. The model parameters are optimized using the Adam algorithm with a learning rate of \(5\times 10^{-5}\) for regularization, a batch size of 6, and a weight decay coefficient of \(L_2\) of \(5\times 10^{-3}\). Training is regularized through early stopping with a patience of 20 epochs, determined by the correlation-based performance metric on a held-out validation set, and all results are averaged over 3 random seeds. 

\begin{table}[h]
    \centering
    \footnotesize
    \setlength{\tabcolsep}{4pt}  % reduce column padding
    \caption{Performance comparison on AVQA benchmarks. All objective scores are mapped to MOS using four-parameter logistic regression~\cite{seshadrinathan2010study}. The best three results per metric are in red, green, and blue, respectively.}
    \label{tab:sota}
    \begin{tabular}{lcccccc}
        \hline
        \multirow{2}{*}{Method} &
        \multicolumn{2}{c}{UnB-AV} &
        \multicolumn{2}{c}{LIVE-SJTU} &
        \multicolumn{2}{c}{UnB-AVQ} \\
        \cline{2-7}
        & PLCC & SROCC & PLCC & SROCC & PLCC & SROCC \\
        \hline
        Linear \cite{martinez2018combining}
        & 0.441 & 0.337 & 0.648 & 0.645 & 0.881 & 0.869 \\
        Minkowski \cite{martinez2018combining}
        & 0.342 & 0.314 & 0.653 & 0.653 & 0.768 & \color{blue}{0.879} \\
        Power \cite{martinez2018combining}
        & 0.662 & 0.608 & 0.628 & 0.640 & 0.887 & 0.862 \\
        NAViDAd~\cite{martinez2019navidad} 
        & \color{blue}{0.881} & \color{green}{0.890} & N/A & N/A & N/A & N/A \\
        DNN-RNT~\cite{cao2023attention}
        & N/A & N/A & \color{green}{0.960} & \color{green}{0.961} & \color{green}{0.904} & \color{green}{0.902} \\
        DNN-SND~\cite{cao2023attention}
        & N/A & N/A & \color{blue}{0.955} & \color{blue}{0.951} & 0.856 & 0.848 \\
        DNFAVQ~\cite{cao2021dnfavq}
        & N/A & N/A & 0.918 & 0.907 & N/A & N/A \\
        Nave+w2v~\cite{martinez2022see}
        & \color{red}{0.936} & \color{red}{0.959} & N/A & N/A & N/A & N/A \\
        UNQA (A/V)~\cite{cao2025unqa} 
        & N/A & N/A & N/A & N/A & \color{blue}{0.903} & 0.863 \\
        \textbf{MCM-AVQA} 
        & \color{green}{0.894} & \color{blue}{0.876} & \color{red}{0.965} & \color{red}{0.970} & \color{red}{0.967} & \color{red}{0.952} \\
        \hline
    \end{tabular}
\end{table}

\begin{table}[h]
    \centering
    \footnotesize
    \setlength{\tabcolsep}{4pt}
    \caption{Statistical significance analysis on UnB-AV.}
    \label{tab:significance_unbav}
    \begin{tabular}{l c}
        \hline
        Quantity & $p$-value \\
        \hline
        Paired $t$-test (two-sided) & $2.1\times10^{-3}$ \\
        Wilcoxon (two-sided) & $4.4\times10^{-3}$ \\
        Wilcoxon (one-sided, MCM-AVQA $<$ Nave+w2v) & $2.2\times10^{-3}$ \\
        \hline
    \end{tabular}
\end{table}

Table~\ref{tab:sota} shows the results of comparing MCM-AVQA to other AVQA approaches, such as classical fusion method \cite{martinez2018combining}, autoencoder-based models~\cite{martinez2019navidad,martinez2022see}, deep learning baselines~\cite{cao2021dnfavq}, and recent attention-driven architectures~\cite{cao2023attention,cao2025unqa}. MCM-AVQA achieves the highest overall performance, as measured by PLCC and SROCC, on both the LIVE-SJTU dataset and the UnB-AVQ dataset, remaining competitive with autoencoder-based algorithms on UnB-AV, Nave+w2v attaining the best correlations. To investigate the per-sequence difference between MCM-AVQA and the Nave+w2v late-fusion baseline on UnB-AV, we employ paired significance tests on absolute prediction errors at a significance level of $\alpha = 0.05$. On the test set of 80 sequences, the difference in the mean absolute error is 0.054 in favor of MCM-AVQA, indicating that MCM-AVQA predictions deviate less from the ground-truth MOS. A paired $t$-test provides a $p$-value of $2.1\times10^{-3}$, while Wilcoxon signed-rank tests provide $p$-values of $4.4\times10^{-3}$ (two-sided) and $2.2\times10^{-3}$ (one-sided, examining if MCM-AVQA has significantly reduced errors), all less than $\alpha$. These findings show that, on UnB-AV, MCM-AVQA achieves much reduced per-sequence absolute error than the Nave+w2v baseline in both parametric and nonparametric analysis.

\section{Ablation Studies}

\begin{table}[t]
    \centering
    \footnotesize
    \caption{Ablation study on UnB-AVQ and LIVE-SJTU. AVM: audio-visual mixer, VCM: visual confidence module, ACM: audio confidence module. AVM (-), VCM (-) and ACM (-) denotes the basic late fusion baseline.}
    \label{tab:ablation}
    \footnotesize
    \setlength{\tabcolsep}{5pt}
    \begin{tabular}{ccc|cc|cc}
        \hline
        \multicolumn{3}{c|}{Configuration} & \multicolumn{2}{c|}{UnB-AVQ} & \multicolumn{2}{c}{LIVE-SJTU} \\
        \hline
        AVM & VCM & ACM & PLCC & SROCC & PLCC & SROCC \\
        \hline
        - & - & - & 0.907 & 0.894 & 0.916 & 0.896 \\
        + & - & - & 0.920 & 0.892 & 0.923 & 0.902 \\
        + & + & - & 0.927 & 0.898 & 0.931 & 0.934 \\
        + & - & + & 0.943 & 0.932 & 0.948 & 0.943 \\
        + & + & + & \textbf{0.967} & \textbf{0.952} & \textbf{0.965} & \textbf{0.970} \\
        \hline
    \end{tabular}
\end{table}

Table~\ref{tab:ablation} shows how different module combinations affect PLCC and SROCC on UnB-AVQ and LIVE-SJTU. The naive late-fusion baseline (AVM-, VCM-, ACM-) has PLCC/SROCC values of 0.907/0.894 on UnB-AVQ and 0.916/0.896 on LIVE-SJTU. Enabling merely the Audio-Visual Mixer without the confidence modules (AVM+, VCM-, and ACM-) improves PLCC to 0.920 on UnB-AVQ and 0.923 on LIVE-SJTU, demonstrating that organized cross-modal fusion is already beneficial. Adding the Visual Confidence Module to AVM without the Audio Confidence Module (AVM+, VCM+, ACM-) improves PLCC to 0.927 and 0.931, and on LIVE-SJTU, it also increases SROCC from 0.902 to 0.934, with a smaller SROCC gain on UnB-AVQ. When all three modules are active (AVM+, VCM+, and ACM+), PLCC achieves 0.967 on UnB-AVQ and 0.965 on LIVE-SJTU, whereas SROCC achieves 0.952 and 0.970, resulting in the best overall performance.

\begin{figure}[tbh]
    \centering
    \includegraphics[width=0.45\textwidth]{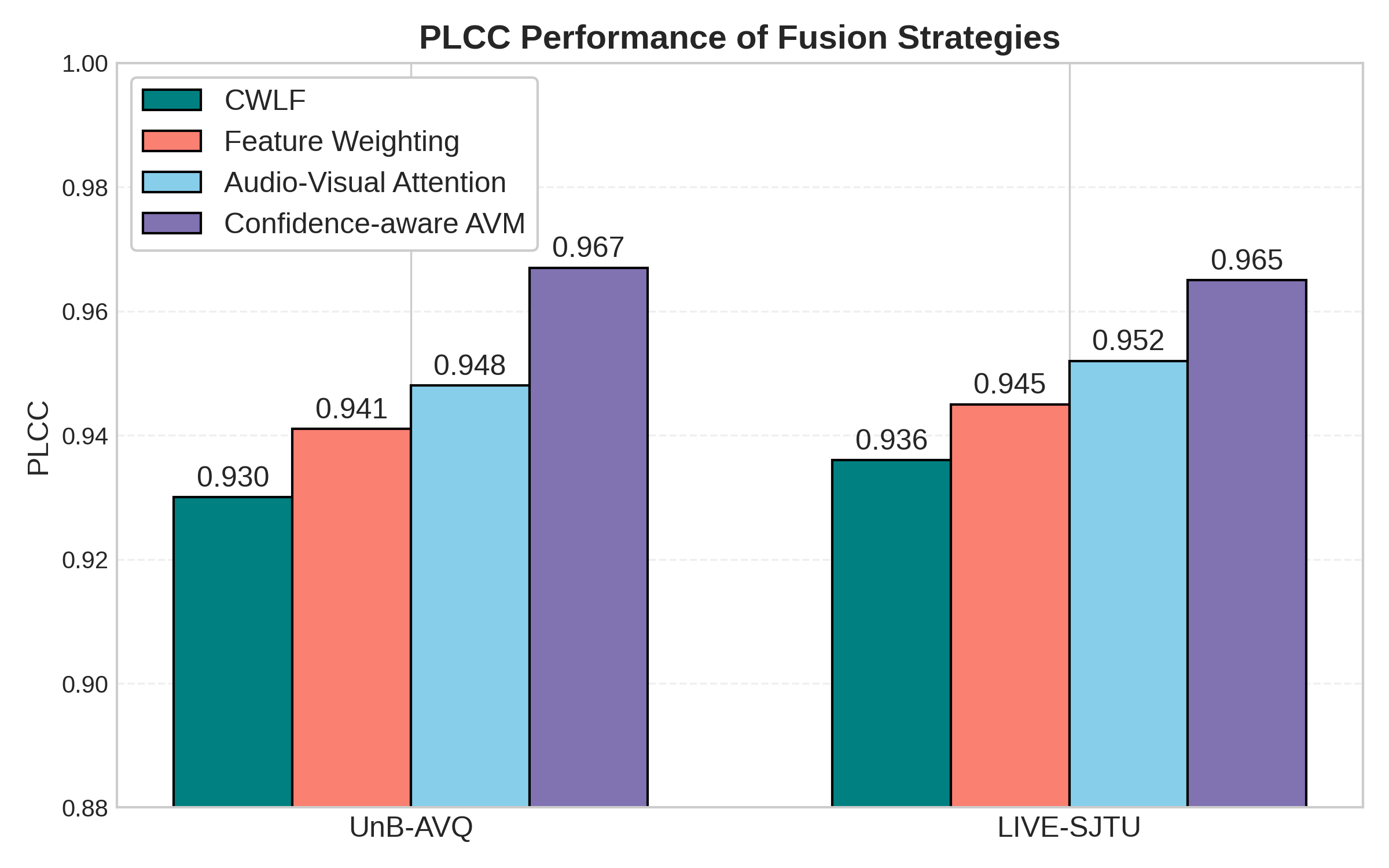}
    \caption{PLCC performance comparison of fusion strategies on UnB-AVQ and LIVE-SJTU, for feature weighting, confidence-weighted late fusion (CWLF), Audio-Visual Attention Network, and our Audio--Visual Mixer (AVM). 
    % Incorporating confidence information improves performance, with AVM achieving the highest PLCC on both datasets.
    }
    \label{fig:fusion_strategies}
\end{figure}

Figure \ref{fig:fusion_strategies} compares 4 fusion strategies on UnB-AVQ and LIVE-SJTU. Confidence-weighted late fusion (CWLF) integrates independent audio-only and visual-only predictions using scalar confidence weights at the decision level, reaching PLCCs of 0.930 and 0.936 but lacking feature-level interaction. Feature weighting increases PLCC to 0.941 and 0.945 by scaling each modality features with clip-level confidence before fusion, yet still relies on a single scalar per modality and ignores channel-specific distortions. The self- and cross-attention Audio-Visual Attention Network~\cite{awan2024attend} further improves PLCC to 0.948 in UnB-AVQ and 0.952 in LIVE-SJTU via Q/K/V-based cross-modal attention, but does not explicitly model modality confidence. Our AVM achieves the best PLCC, 0.967 on UnB-AVQ and 0.965 on LIVE-SJTU, by embedding modality confidence into channel-specific cross-modal attention, allowing audio and visual confidences to directly control which channels are emphasized or suppressed.

\begin{figure}[t]
    \centering
    \includegraphics[width=0.45\textwidth]{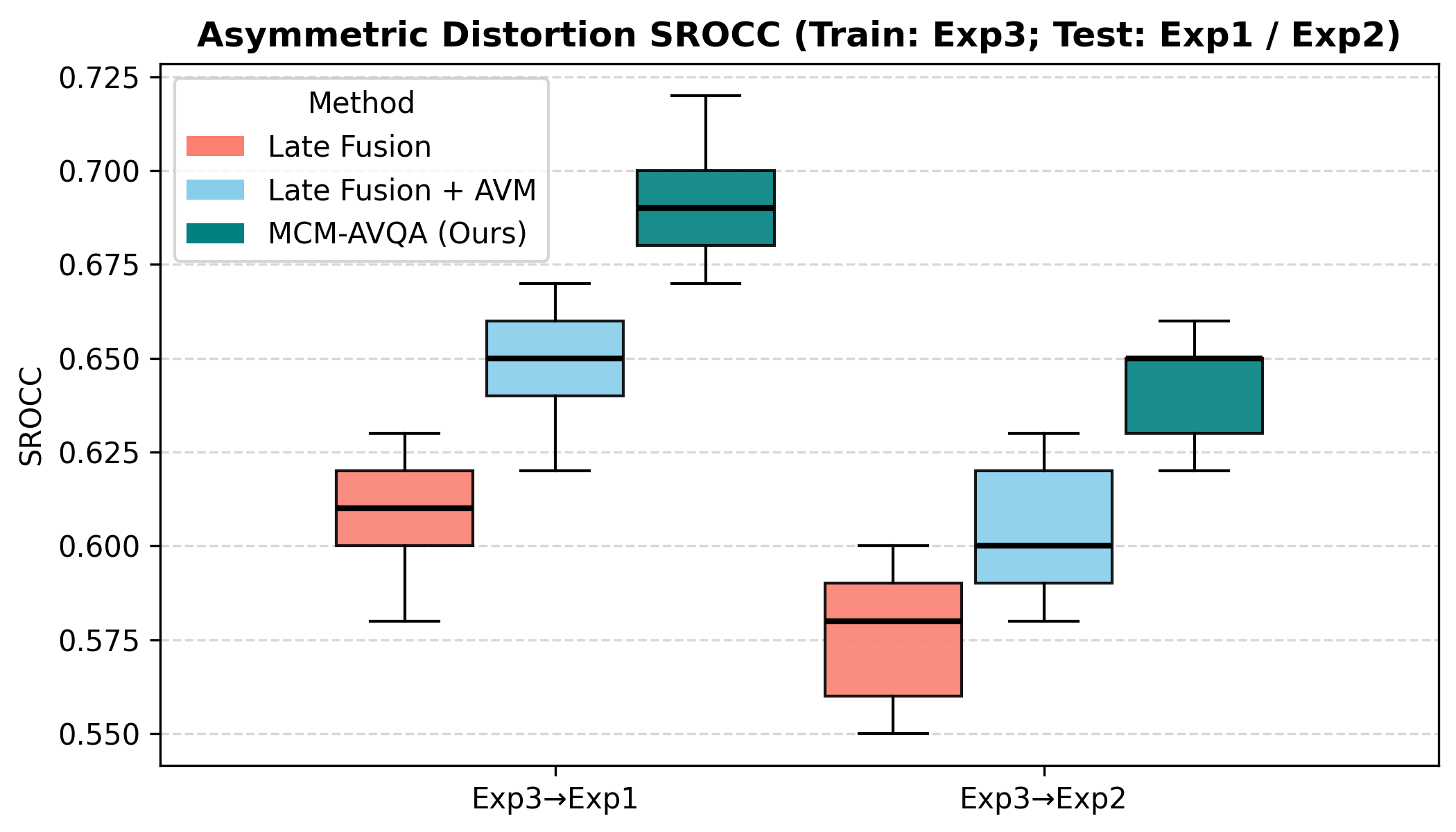}
    \caption{Asymmetric distortion analysis on UnB-AV. Models are trained on Exp3 and evaluated on Exp1 (video-degraded only) and Exp2 (audio-degraded only). Each box plot summarizes the SROCC distribution across five runs, where the central black line denotes the median.}
    \label{fig:unbav_asymmetric_srocc}
\end{figure}

Figure~\ref{fig:unbav_asymmetric_srocc} assesses adaptability to asymmetric distortions using UnB-AV Exp1 (video distorted, audio clean) and Exp2 (audio degraded, video clean). All models are trained in Exp3, then tested in Exp1 and Exp2 over five runs. Late fusion has the lowest median SROCC and the widest boxes in both situations, indicating that it performs poorly when one modality is significantly worse than the other. Adding the Audio-Visual Mixer (Late Fusion + AVM) raises the median SROCC to around 0.65 in Exp1 and 0.60 in Exp2, while also reducing variability, indicating that structured cross-modal fusion is already beneficial. MCM-AVQA achieves the greatest median SROCC in both settings, approximately 0.70 for Exp3→Exp1 and 0.65 for Exp3→Exp2, with the tightest boxes among all methods. When audio degradation occurs (Exp2), performance decreases for all models. However, MCM-AVQA has a smaller drop from Exp3→Exp1 to Exp3→Exp2, indicating that confidence-aware fusion is more robust to asymmetric distortions.

\section{Conclusion}
This study presents MCM-AVQA, a confidence-aware audio-visual quality assessment framework that first models modality-specific confidence, then feeds it into an Audio-Visual Mixer for cross-modal integration. MCM-AVQA adapts to asymmetric distortion, where one modality is heavily degraded and the other remains reliable. Experiments on LIVE-SJTU, UnB-AV, and UnB-AVQ show that MCM-AVQA consistently outperforms state-of-the-art methods in PLCC and SROCC across varied distortions. Ablation studies indicate that the Audio-Visual Mixer and explicit modality confidence estimation are crucial for improving adaptability and temporal stability, particularly under highly asymmetric degradation. Future work will extend confidence modeling to more real-world audiovisual distortions and explore its use in other multimodal quality assessment tasks.

\bibliographystyle{IEEEbib}
\bibliography{refs}

\begin{thebibliography}{10}

\bibitem{akhtar2017audio}
Zahid Akhtar and Tiago~H Falk,
\newblock ``Audio-visual multimedia quality assessment: A comprehensive
  survey,''
\newblock {\em IEEE access}, vol. 5, pp. 21090--21117, 2017.

\bibitem{min2020livesjtu}
Xiongkuo Min, Guangtao Zhai, Jiantao Zhou, Mylene~CQ Farias, and Alan~Conrad
  Bovik,
\newblock ``Study of subjective and objective quality assessment of
  audio-visual signals,''
\newblock {\em IEEE Transactions on Image Processing}, vol. 29, pp. 6054--6068,
  2020.

\bibitem{rohe2018reliability}
Tim Rohe and Uta Noppeney,
\newblock ``Reliability-weighted integration of audiovisual signals can be
  modulated by top-down attention,''
\newblock {\em eneuro}, vol. 5, no. 1, 2018.

\bibitem{cao2023attention}
Yuqin Cao, Xiongkuo Min, Wei Sun, and Guangtao Zhai,
\newblock ``Attention-guided neural networks for full-reference and
  no-reference audio-visual quality assessment,''
\newblock {\em IEEE Transactions on Image Processing}, vol. 32, pp. 1882--1896,
  2023.

\bibitem{martinez2019navidad}
Helard Martinez, Myl{\`e}ne C.~Q. Farias, and Andrew Hines,
\newblock ``{NAViDAd}: A no-reference audio-visual quality metric based on a
  deep autoencoder,''
\newblock in {\em Proc. EUSIPCO}. 2019, IEEE.

\bibitem{gao2024avsegformer}
Shengyi Gao, Zhe Chen, Guo Chen, Wenhai Wang, and Tong Lu,
\newblock ``Avsegformer: Audio-visual segmentation with transformer,''
\newblock in {\em Proceedings of the AAAI conference on artificial
  intelligence}, 2024, vol.~38, pp. 12155--12163.

\bibitem{cheng2025mmaudio}
Ho~Kei Cheng, Masato Ishii, Akio Hayakawa, Takashi Shibuya, Alexander Schwing,
  and Yuki Mitsufuji,
\newblock ``Mmaudio: Taming multimodal joint training for high-quality
  video-to-audio synthesis,''
\newblock in {\em Proceedings of the Computer Vision and Pattern Recognition
  Conference}, 2025, pp. 28901--28911.

\bibitem{mithila2026convolutions}
Mayesha Maliha~Rahman Mithila and Mylene~CQ Farias,
\newblock ``Convolutions need registers too: Hvs-inspired dynamic attention for
  video quality assessment,''
\newblock in {\em Proceedings of the ACM Multimedia Systems Conference 2026},
  2026, pp. 37--48.

\bibitem{chakraborty2025mt}
Swarna Chakraborty and Mylene~CQ Farias,
\newblock ``Mt-dpcqa: A multimodal time-aware learning approach for
  no-reference dynamic point cloud quality assessment,''
\newblock in {\em Proceedings of the 33rd ACM International Conference on
  Multimedia}, 2025, pp. 7113--7122.

\bibitem{awan2024attend}
Mahrukh Awan, Asmar Nadeem, Muhammad~Junaid Awan, Armin Mustafa, and
  Syed~Sameed Husain,
\newblock ``Attend-fusion: Efficient audio-visual fusion for video
  classification,''
\newblock in {\em European Conference on Computer Vision}. Springer, 2024, pp.
  195--213.

\bibitem{cao2025unqa}
Yuqin Cao, Xiongkuo Min, Yixuan Gao, Wei Sun, Long Ye, Weisi Lin, and Guangtao
  Zhai,
\newblock ``Unqa: Unified no-reference quality assessment for audio, image,
  video, and audio-visual content,''
\newblock {\em IEEE Transactions on Circuits and Systems for Video Technology},
  2025.

\bibitem{alam2015confidence}
Mohammad~Rafiqul Alam, Mohammed Bennamoun, Roberto Togneri, and Ferdous Sohel,
\newblock ``A confidence-based late fusion framework for audio-visual biometric
  identification,''
\newblock {\em Pattern Recognition Letters}, vol. 52, pp. 65--71, 2015.

\bibitem{feng2025mvad}
Chen Feng, Duolikun Danier, Fan Zhang, Alex Mackin, Andrew Collins, and David
  Bull,
\newblock ``Mvad: A multiple visual artifact detector for video streaming,''
\newblock in {\em 2025 IEEE/CVF Winter Conference on Applications of Computer
  Vision (WACV)}. IEEE, 2025, pp. 3148--3158.

\bibitem{ragano2024scoreq}
Alessandro Ragano, Jan Skoglund, and Andrew Hines,
\newblock ``Scoreq: Speech quality assessment with contrastive regression,''
\newblock {\em Advances in Neural Information Processing Systems}, vol. 37, pp.
  105702--105729, 2024.

\bibitem{liu2021swin}
Ze~Liu, Yutong Lin, Yue Cao, Han Hu, Yixuan Wei, Zheng Zhang, Stephen Lin, and
  Baining Guo,
\newblock ``Swin transformer: Hierarchical vision transformer using shifted
  windows,''
\newblock in {\em Proceedings of the IEEE/CVF international conference on
  computer vision}, 2021, pp. 10012--10022.

\bibitem{hershey2017cnn}
Shawn Hershey, Sourish Chaudhuri, Daniel~PW Ellis, Jort~F Gemmeke, Aren Jansen,
  R~Channing Moore, Manoj Plakal, Devin Platt, Rif~A Saurous, Bryan Seybold,
  et~al.,
\newblock ``Cnn architectures for large-scale audio classification,''
\newblock in {\em 2017 ieee international conference on acoustics, speech and
  signal processing (icassp)}. IEEE, 2017, pp. 131--135.

\bibitem{martinez2020unbav}
Helard~B. Martinez, Andrew Hines, and Myl{\`e}ne C.~Q. Farias,
\newblock ``{UnB-AV}: An audio-visual database for multimedia quality
  research,''
\newblock {\em IEEE Access}, vol. 8, pp. 56641--56649, 2020.

\bibitem{martinez2014full}
Helard~B. Martinez and Myl{\`e}ne~CQ Farias,
\newblock ``Full-reference audio-visual video quality metric,''
\newblock {\em Journal of Electronic Imaging}, vol. 23, no. 6, pp.
  061108--061108, 2014.

\bibitem{seshadrinathan2010study}
Kalpana Seshadrinathan, Rajiv Soundararajan, Alan~Conrad Bovik, and Lawrence~K
  Cormack,
\newblock ``Study of subjective and objective quality assessment of video,''
\newblock {\em IEEE transactions on Image Processing}, vol. 19, no. 6, pp.
  1427--1441, 2010.

\bibitem{martinez2018combining}
Helard~Becerra Martinez and Myl{\`e}ne C.~Q. Farias,
\newblock ``Combining audio and video metrics to assess audio-visual quality,''
\newblock {\em Multimedia Tools and Applications}, vol. 77, no. 21, pp.
  28449--28474, 2018.

\bibitem{cao2021dnfavq}
Yuqin Cao, Xiongkuo Min, Wenhan Sun, and Guangtao Zhai,
\newblock ``Deep neural networks for full-reference and no-reference
  audio-visual quality assessment,''
\newblock in {\em Proceedings of the IEEE International Conference on Image
  Processing (ICIP)}, 2021, pp. 1429--1433.

\bibitem{martinez2022see}
Helard~B Martinez, Andrew Hines, and Myl{\`e}ne~CQ Farias,
\newblock ``See hear now: is audio-visual qoe now just a fusion of audio and
  video metrics?,''
\newblock in {\em 2022 14th International Conference on Quality of Multimedia
  Experience (QoMEX)}. IEEE, 2022, pp. 1--4.

\end{thebibliography}

\end{document}